\documentclass{article}

   \usepackage[final,nonatbib]{neurips_2024}

\usepackage[utf8]{inputenc} 
\usepackage[T1]{fontenc}    
\usepackage{hyperref}       
\usepackage{url}            
\usepackage{booktabs}       
\usepackage{amsfonts}       
\usepackage{nicefrac}       
\usepackage{microtype}      
\usepackage{xcolor}         
\usepackage{graphicx}
\usepackage{enumitem}
\usepackage{subcaption}
\usepackage{bm}
\usepackage{multirow}
\usepackage[numbers]{natbib}
\title{DARNet: Dual Attention Refinement Network with Spatiotemporal Construction for Auditory Attention Detection}

\author{
  Sheng Yan\(^{1}\)\thanks{Equal contribution.} 
  \And Cunhang Fan\(^{1*}\)\thanks{Corresponding author. Correspondence to cunhang.fan@ahu.edu.cn.} 
  \And Hongyu Zhang\(^{1}\)
  \And Xiaoke Yang\(^{1}\)
  \And Jianhua Tao\(^{2}\)
  \And Zhao Lv\(^{1}\)
  \\ \(^{1}\)Anhui Province Key Laboratory of Multimodal Cognitive Computations,
    \\ School of Computer Science and Technology, Anhui University
    \\\(^{2}\)Department of Automation, Tsinghua University 
}

\begin{document}

\maketitle

\begin{abstract}
At a cocktail party, humans exhibit an impressive ability to direct their attention. The auditory attention detection (AAD) approach seeks to identify the attended speaker by analyzing brain signals, such as EEG signals. 
However, current AAD algorithms overlook the spatial distribution information within EEG signals and lack the ability to capture long-range latent dependencies, limiting the model's ability to decode brain activity.
To address these issues, this paper proposes a dual attention refinement network with spatiotemporal construction for AAD, named DARNet, which consists of the spatiotemporal construction module, dual attention refinement module, and feature fusion \& classifier module. Specifically, the spatiotemporal construction module aims to construct more expressive spatiotemporal feature representations, by capturing the spatial distribution characteristics of EEG signals. The dual attention refinement module aims to extract different levels of temporal patterns in EEG signals and enhance the model's ability to capture long-range latent dependencies. The feature fusion \& classifier module aims to aggregate temporal patterns and dependencies from different levels and obtain the final classification results.
The experimental results indicate that DARNet achieved excellent classification performance, particularly under short decision windows. While maintaining excellent classification performance, DARNet significantly reduces the number of required parameters. Compared to the state-of-the-art models, DARNet reduces the parameter count by 91\%. Code is available at: \href{https://github.com/fchest/DARNet.git}{https://github.com/fchest/DARNet.git}.
\end{abstract}

\section{Introduction}

The auditory attention detection (AAD) aims to study human auditory attention tendencies by analyzing brain signals \citep{4,monesi2020lstm,3}. The auditory attention refers to the ability of individuals to isolate or concentrate on specific sounds, which aids them in focusing on a single speaker amidst a multi-speaker environment, a scenario commonly referred to as the "cocktail party scenario" \citep{1}. However, this ability may diminish or even completely disappear for individuals with impairment. Therefore, finding solutions to assist these individuals in overcoming this challenge has become an urgent matter. 

\citet{5} have demonstrated a close connection between auditory attention and brain activity, which indicates that researchers can study auditory attention by analyzing brain activity. Following this concept, many methods such as electrocorticography (ECoG) \citep{5}, magnetoencephalography \citep{8,9}, and electroencephalography (EEG) \citep{6,7} have been used to implement auditory attention detection. Among these methods, EEG-based approaches are widely applied in AAD due to their high temporal resolution, non-invasive mode, and excellent maneuverability \citep{7,10,11}.

According to the conclusions of \citet{5}, previous studies have utilized stimulus-reconstruction or speech envelope reconstruction methods, which necessitate clean auditory stimuli as input \citep{aroudi2018eeg,das2017eeg}. However, in most real-world scenarios, environments consist of multiple sounds simultaneously. Listeners are exposed to a mixture of these sounds, posing a challenge in obtaining clean auditory stimuli. Therefore, in recent years, the academic community has increasingly focused solely on utilizing EEG signals as input for AAD research \citep{stanet,mbssf, fan2024dgsd}. The research method proposed in this paper also exclusively utilizes EEG signals.

Traditional AAD tasks relied on linear models to process EEG signals \citep{crosse2016multivariate,wong2018comparison}. However, brain activity is inherently nonlinear, posing challenges for linear models in capturing this complexity. Consequently, they necessitate longer decision windows to extract brain activity features \citep{miran2018real}. 
Some previous studies have indicated that decent decoding performance can be achieved by analyzing different spatial distribution features within each frequency band. These studies project the extracted differential entropy (DE) values onto 2D topological maps and decode them with convolutional neural networks \citep{DBPNet, mbssf}. However, EEG signals are fundamentally time-series data, these methods overlook the dynamic temporal patterns of EEG signals. Other studies analyze EEG signals only in the time domain. For instance, they use long short-term memory (LSTM) networks to capture dependencies within EEG signals and achieve decent decoding performance \citep{monesi2020lstm}. However, these studies only focus on the temporal information within EEG signals, neglecting the spatial distribution features, which reflect the dynamic patterns of different brain regions when receiving, processing, and responding to auditory stimuli. Meanwhile, numerous noise points and outliers make it difficult to capture long-range latent dependencies.

To address these issues, this paper proposes a dual attention refinement network with spatiotemporal construction for AAD, named DARNet, which effectively captures the spatiotemporal features and long-range latent dependencies of EEG signals. Specifically, our model consists of three modules: (1) \textit{Spatiotemporal Construction Module}. The spatiotemporal construction module employs a temporal convolutional layer and a spatial convolutional layer. The temporal convolutional layer effectively captures the temporal dynamic features of EEG signals, and the spatial convolutional layer captures the spatial distribution features among different channels, thereby constructing a robust embedding for the next layer. (2) \textit{Dual Attention Refinement Module}. The dual-layer self-attention refinement module consists of two layers, each comprising a multi-head self-attention and a refinement layer. This design is intended to capture long-range latent dependencies and deeper sequence patterns in EEG signals. (3) \textit{Feature Fusion \& Classifier Module}. The attention features generated by the dual-layer self-attention refinement module, comprising both shallow and deep levels, are fed into the feature fusion module to obtain richer representations, enhancing the model's robustness and generalization. The fused features are input into a classifier to predict the auditory attention tendencies of the subjects.

To this end, We evaluated the decoding performance of DARNet on three datasets: DTU, KUL, and MM-AAD. The results demonstrate that DARNet outperforms the current state-of-the-art model on all three datasets. The main contributions of this paper are summarized as follows: 
\begin{itemize}
    \item We propose a novel auditory attention decoding architecture, which consists of a spatiotemporal construction module, a dual attention refinement module, and a feature fusion module. This architecture could fully leverage the spatiotemporal features and capture long-range latent dependencies of EEG signals.
    \item The DARNet achieves remarkable decoding accuracy within very short decision windows, surpassing the current state-of-the-art (SOTA) model by 5.9\% on the DTU dataset and 4.9\% on the KUL dataset, all under a 0.1-second decision window. Furthermore, compared to the current state-of-the-art model with 0.91 million training parameters, DARNet achieves further parameter reduction, requiring only 0.08 million parameters.
\end{itemize}
\newpage

\begin{figure}[ht]
    \centering
    \includegraphics[width=1\linewidth]{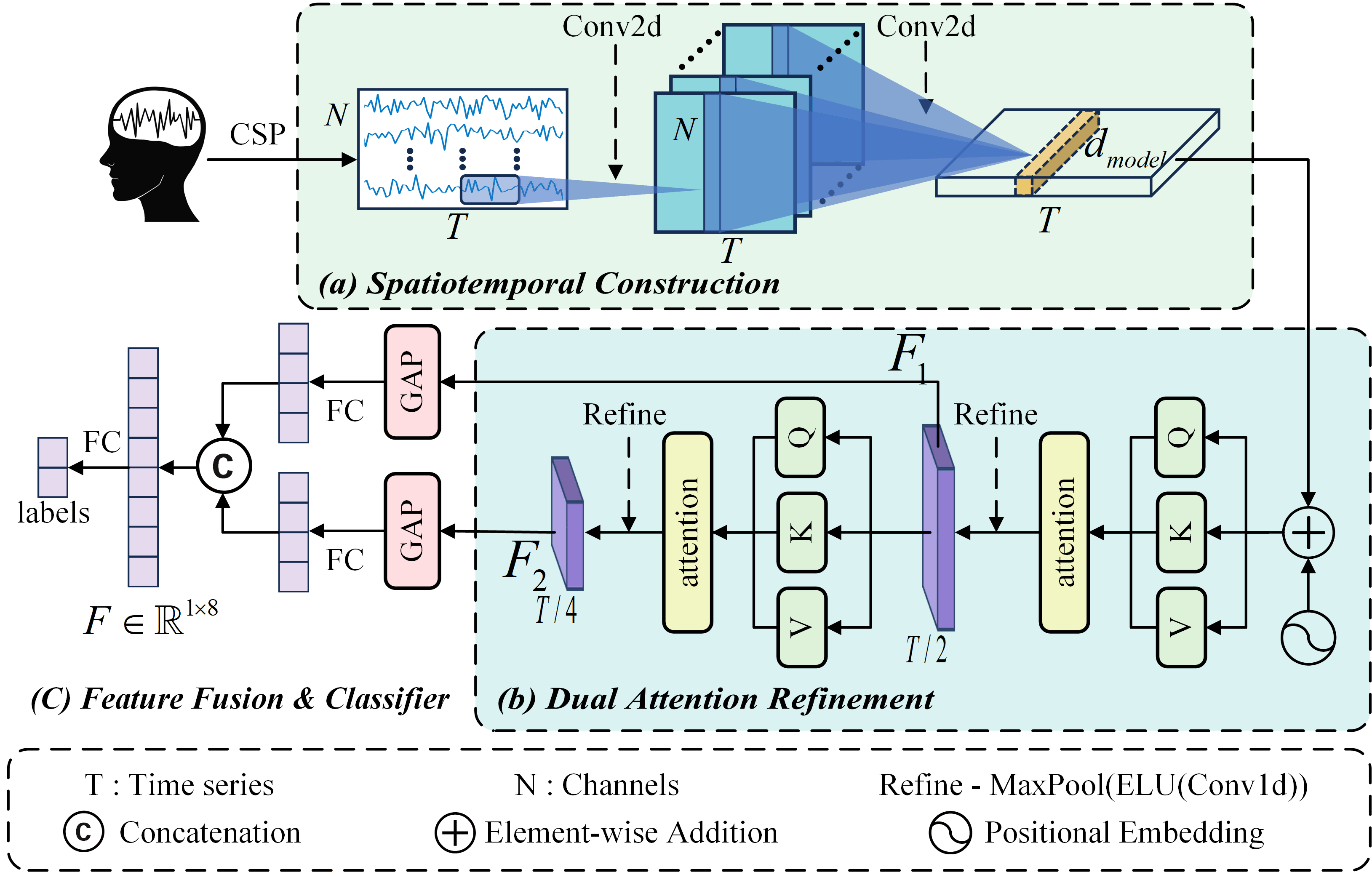}
    \caption{The framework of the DARNet model for AAD, which mainly consists of three modules: (a) spatiotemporal construction module, (b) dual attention refinement module, and (c) feature fusion \& classifier module. The model inputs are common spatial patterns (CSP) extracted from EEG signals, and the outputs are two predicted labels related to auditory attention.}
    \label{model-figure}
\end{figure}
\section{Methodology}
The previous AAD methods overlooked the influence of spatial distribution characteristics on decoding performance and struggled to capture the long-range dependencies in EEG signals \citep{stanet, DBPNet}. To address these issues, we proposed DARNet, which consists of a spatiotemporal construction module, a dual attention refinement module, and a feature fusion \& classifier module, see Figure~\ref{model-figure}. Our proposed DARNet effectively captures the spatiotemporal features of EEG signals and has the capability to capture long-range latent dependencies in EEG signals.

By employing a moving window on the EEG data, we obtain a series of decision windows, each containing a small duration of EEG signals. Let \( R = [r_1,...,r_i,...,r_N] \in \mathbb{R}^{T \times N} \) represents the EEG signals of a decision window, where \(r_i \in \mathbb{R}^{N \times 1}\) represents the EEG data at the \(i\)-th time point within a decision window, contains \(N\) channels. Here \(N\) represents the number of EEG channels and \(T\) denotes the length of the decision window. Before inputting EEG data into the DARNet, we employ a common spatial patterns (CSP) algorithm to extract raw features from the EEG data under different brain states \citep{ramoser2000optimal,blankertz2007optimizing}. 
\begin{equation}
    E = CSP(R) \in \mathbb{R}^{c\_in \times T}
\end{equation}
where \(CSP(\cdot)\) represents the CSP algorithm, \( E \in \mathbb{R}^{N\times T}\) represents the processed EEG signal. \(c\_in\) is the components of the CSP algorithm and T denotes the length of the decision window.

\subsection{Spatiotemporal Construction Module}
EEG signals record the brain's neuronal electrical activity, varying over time and reflecting activity patterns and connectivity across brain regions \citep{arvaneh2011optimizing}. By constructing spatiotemporal features from EEG signals, it's possible to analyze the brain's response patterns to auditory stimuli. However, previous studies only focused on local temporal patterns in EEG data, overlooking the spatial distribution features. Therefore, in addition to the conventional use of temporal filters, we introduced a spatial filter \citep{foumani2024improving} to construct the spatiotemporal features of EEG signals.

Firstly, we use temporal convolution layers to capture the instantaneous changes in EEG signals, thereby constructing the temporal patterns \(E_t\) of the EEG signals. It can be formulated as follows:
\begin{equation}
    E_t = GELU(TemporalConv2d(E)) \in \mathbb{R}^{4d_{model} \times c\_in \times T}
\end{equation}
where \(TemporalConv2d(\cdot)\) performs an 2-D convolutional filters (kernel size=\(1 \times 8\)) on time dimension with \(GELU(\cdot)\) activation function. \(d_{model}\) represents the embedding dimension.

Subsequently, we employ a spatial convolutional layer with a receptive field spanning all channels to capture the spatial distribution features \(S\) of EEG signals across different channels, thereby aiding the model in comprehensively understanding the brain's activity patterns in response to various auditory stimuli. 
\begin{equation}
    S = GELU(SpatialConv2d(E_t)) \in \mathbb{R}^{d_{model} \times T}
\end{equation}
 where \(SpatialConv2d(\cdot)\) performs an 2-D convolutional filters with a \(c\_in \times 1\) kernel size on spatial dimension. By doing so, we not only capture the temporal patterns in EEG signals but also integrate the spatial distribution characteristics of EEG signals, thereby constructing input embedding \(S\) containing comprehensive spatiotemporal information for the next layer. This integrated input better reflects the complex features within EEG signals, providing richer information for subsequent analysis and processing. 
 
\subsection{Dual Attention Refinement Module}
Previous psycho-acoustic research has demonstrated that human attention is a dynamic and time-related activity \citep{jones1989dynamic,jones2002temporal}. The brain activity from the preceding moment can profoundly influence subsequent brain activity \citep{linkenkaer2001long}. However, previous AAD algorithms were hindered by model depth and the noise and outliers in EEG data, making them ineffective at capturing the long-range latent dependencies in EEG signals. 

To address this issue, we proposed a dual self-attention mechanism, which has greater potential for capturing long-range latent dependencies and deeper sequence patterns in EEG signals. Inspired by \citet{zhou2021informer,yu2017dilated}, We introduced a self-attention refinement operation, which refines the dominant temporal features through convolution and pooling operations, compressing the original EEG series of length \(T\) to half its length. This self-attention refinement operation reduces the impact of noise and outliers, while also decreasing the model's parameter count. This enhances the model's generalization and robustness. The single-layer attention refinement module can be formulated as follows:
\begin{equation}
    F = MaxPool(ELU(Conv1d(MultiHeadAttention(x))))
\end{equation}
where \(MultiHeadAttention(\cdot)\) denotes multi-head self-attention algorithm \citep{vaswani2017attention}, \(Conv1d(\cdot)\) represents an 1-D convolutional filters (kernel width=3) on time dimension. The \(ELU(\cdot)\) is the activation function proposed by \citet{clevert2015fast}, \(MaxPool(\cdot)\) denotes a max-pooling layer with stride \(2\). 

Before applying the temporal attention feature extraction module, we add the absolute positional embedding \citep{vaswani2017attention} to the input embedding \(S\) as follows:
\begin{equation}
    s_i = s_i + p_i
\end{equation}
where \(s_i\) represents the embedding vector of the \(i_{th}\) time step, \(p_i \in \mathbb{R}^{d_{model}}\) represents \(i_{th}\) time step position.

To obtain different levels of temporal features from EEG signals and to capture the long-range latent dependencies, we stacked two of the above attention refinement extraction modules.
\begin{equation}
    F_1 = MaxPool(ELU(Conv1d(MultiHeadAttention(S)))) \in \mathbb{R}^{d_{model} \times \frac{T}{2}}
\end{equation}
\begin{equation}
    F_2 = MaxPool(ELU(Conv1d(MultiHeadAttention(F_1)))) \in \mathbb{R}^{d_{model} \times \frac{T}{4}}
\end{equation}
where \(F_1\) and \(F_2\) contain different levels of dependencies and temporal patterns in the EEG signals, respectively.

\subsection{Feature Fusion \& Classifier Module}
Features at different levels can reflect various characteristics of the pattern. By optimizing and combining these different features, it preserves effective discriminative information from features at different levels while also to some extent eliminating redundant information \citep{sun2005new}. Therefore, we designed a feature fusion module as follows:

First, we project the features \(F_1\) and \(F_2\) into the same dimension.
\begin{equation}
    F_1' = Linear(AdaptiveAvgPool(F_1)) \in \mathbb{R}^4
\end{equation}
\begin{equation}
    F_2' = Linear(AdaptiveAvgPool(F_2)) \in \mathbb{R}^4
\end{equation}
where \(AdaptiveAvgPool(\cdot)\) denotes an adaptive average pooling layer, \(Linear\) denotes a linear layer.

Second, We concatenate feature \(F_1'\) and \(F_2'\) to obtain the fused feature vector \(F\).
\begin{equation}
    F = [F_1', F_2']
\end{equation}

Finally, we employ a fully connected layer to obtain the final auditory attention prediction.
\begin{equation}
    p = w(F + b)
\end{equation}
where \(w\) and \(b\) are the weight and the bias of the fully connected layer, \(p\) denotes the predicted direction label. In the training stage, we employ the cross entropy loss function to supervise the network training.

\section{Experiments}
\subsection{Dataset}
In this section, we conduct experiments on three publicly available datasets, namely KUL \citep{das2016effect,das2019auditory}, DTU \citep{fuglsang2017noise,fuglsang2018eeg} and MM-AAD \citep{DBPNet}, which are commonly used in auditory attention detection to evaluate the effectiveness of our DARNet. KUL and DTU only contain EEG data of the auditory stimulus scenes. MM-AAD contains EEG data of the audio-only scene and the audio-visual scene. We summarize the details of the above datasets in Table~\ref{dataset-table}.
\begin{enumerate}
    \renewcommand{\labelenumi}{\arabic{enumi})}
    \item \textbf{KUL Dataset:} In this dataset, 64-channel EEG data were collected from 16 normal-hearing subjects using a BioSemi ActiveTwo device at a sampling rate of 8,192 Hz in a soundproof room. Each subject was instructed to focus on one of two simultaneous speakers. The auditory stimuli were filtered at 4kHz and set at 60dB through in-ear headphones, which contain four Dutch short stories, narrated by three male Flemish speakers. Two listening conditions were employed: dichotic (dry) presentation with one speaker per ear, and head-related transfer function (HRTF) filtered presentation, simulating speech from 90° left or right. Each subject listened to 8 trials, which lasted 6 minutes.
    
    \item \textbf{DTU Dataset:} In this dataset, 64-channel EEG data were collected from 18 normal-hearing subjects using a BioSemi ActiveTwo device at a sampling rate of 512 Hz. Each subject was instructed to focus on one of two simultaneous speakers, who presented at 60° relative to the subject. The auditory stimuli were set at 60dB through ER-2 earphones, which contain Danish audiobooks, narrated by three male speakers and three female speakers. Each subject listened to 60 trials, which lasted 50 seconds.
    
    \item \textbf{MM-AAD Dataset:} In this dataset, 32-channel EEG data were collected from 50 normal-hearing subjects (34 males and 16 females) at a sampling rate of 4kHz, following the 10/20 international system. Each subject was exposed to both audio-only and audio-visual stimuli. They were instructed to focus on one of two simultaneous speakers, who presented at left or right spatial direction relative to the subject. The auditory stimuli comprised 40 classic Chinese stories narrated by both male and female speakers. Each subject listened to 20 trials, which lasted 165 seconds.
\end{enumerate}

\begin{table}
    \caption{Details of three datasets used in the experiments.}
    \label{dataset-table}
    \centering
    \begin{tabular}{cccccc}
        \toprule
        Dataset & Subjects & Scene & Language & \parbox{3cm}{\centering Duration per subject (minutes)} & \parbox{2cm}{\centering Total duration (hours)} \\
        \midrule
        KUL & 16 & audio-only & Dutch & 48 & 12.8 \\
        DTU & 18 & audio-only & Danish & 50 & 15.0\\
        \multirow{2}{*}{MM-AAD} & 50 & audio-only & Chinese & 55 & 45.8 \\
         & 50 & audio-visual & Chinese & 55 & 45.8 \\
        \bottomrule
    \end{tabular}
\end{table}

\subsection{Data Processing}
To fairly compare the performance of the proposed DARNet model, specific preprocessing steps are applied to each dataset (KUL, DTU, and MM-AAD). For the KUL dataset, the EEG data were firstly re-referenced to the average response of mastoid electrodes, then bandpass filtered between 0.1 Hz and 50 Hz, and finally down-sampled to 128 Hz. For the DTU dataset, the EEG data were filtered to remove 50 Hz linear noise and harmonics. Eye artifacts were eliminated through joint decorrelation and the EEG data were re-referenced to the average response of mastoid electrodes. Finally, the EEG data were down-sampled to 64 Hz. For the MM-AAD dataset, the EEG data were firstly bandpass filtered between 0.1 Hz and 50 Hz, then removed 50 Hz noise through a notch filter. Additionally, eye artifacts were eliminated, and further noise removal was achieved, using independent component analysis (ICA). Finally, the EEG data were down-sampled to 128 Hz.

We evaluated our proposed DARNet model and compared it with other state-of-the-art models under three decision window lengths: 0.1s, 1s, and 2s. Specifically, we selected three publicly available models as our baseline for comparison: SSF-CNN \citep{cai2021low}, MBSSFCC \citep{mbssf}, and DBPNet \citep{DBPNet}.

\subsection{Implement Details}
In previous AAD research, the accuracy of auditory attention prediction classification has been used as a benchmark for model performance. We followed this convention and evaluated our proposed DARNet on the KUL, DTU, and MM-AAD datasets. As follows, we take the KUL dataset with a 1-second decision window as an example to illustrate implementation details, including training settings and network configuration.

Firstly, we set the proportions of the training, validation, and test sets to 8:1:1. For each subject of the KUL dataset, we get 4,600 decision windows for training, 576 decision windows for validation, and 576 decision windows for testing. Meanwhile, we set the batch size to 32, the maximum number of epochs to 100, and employ an early stopping strategy. Training will stop if the loss function value on the validation set does not decrease for 10 consecutive epochs. Additionally, we utilize the Adam optimizer with a learning rate of 5e-4 and weight decay of 3e-4 to train the model. The DARNet is performed using PyTorch.

Before inputting EEG data into the DARNet, we employ the CSP algorithm to extract raw features \(E \in \mathbb{R}^{128 \times 64}\) from the EEG data. The data is transposed and expanded, represented as \(E' \in \mathbb{R}^{1 \times 64 \times 128}\). Then, through the spatiotemporal construction module (\(c_{in}\) is set to 16), we can get embedding data \(S \in \mathbb{R}^{16 \times 1 \times 128}\). After dimensionality reduction, transposition, and the addition of absolute positional embedding, the data is fed into the dual attention refinement module, resulting in two distinct level features, \(F_1 \in \mathbb{R}^{16 \times 64 }\) and \(F_2 \in \mathbb{R}^{16 \times 32} \). The \(F_1\) and \(F_2\) are sent to the feature fusion module, where they undergo global average pooling and dimensionality reduction via a fully connected (FC) layer (input: 16, output: 4) before being concatenated to obtain the fused feature, \(F \in \mathbb{R}^{8} \). Finally, \(F\) is passed through another FC layer (input: 8, output:2) to obtain the final auditory attention prediction \(p \in \mathbb{R}^2\).

\section{Result}
\subsection{Performance of DARNet}
\begin{figure}[ht]
    \centering
    \begin{minipage}[t]{0.45\textwidth}
        \centering
        \includegraphics[width=\linewidth]{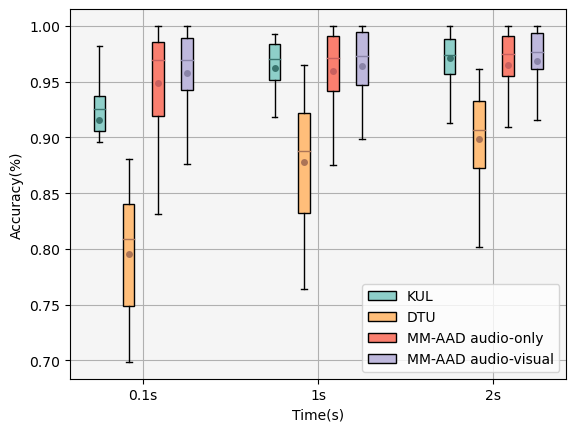}
        \caption{AAD accuracy(\%) of DARNet across all subjects on three datasets.}
        \label{compare_figure}
    \end{minipage}
    \hspace{0.05\textwidth}
    \begin{minipage}[t]{0.45\textwidth}
        \centering
        \includegraphics[width=\linewidth]{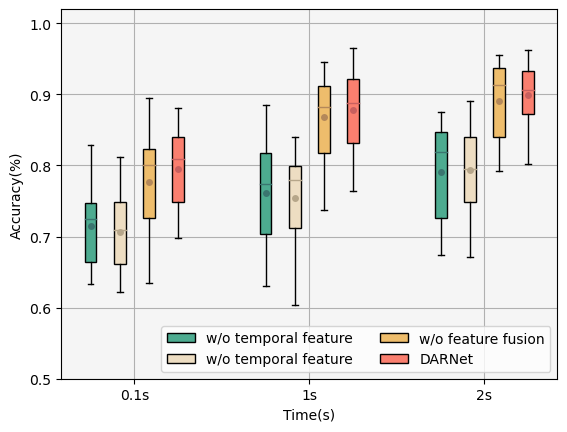}
        \caption{AAD accuracy(\%) of the ablation study across all subjects on the DTU dataset.}
        \label{ablation_figure}
    \end{minipage}
\end{figure}

To evaluate the performance of DARNet, we conducted comprehensive experiments under decision windows of 0.1-second, 1-second, and 2-second, respectively, as shown in Figure~\ref{compare_figure}. Additionally, We compared our DARNet with other advanced models, as shown in Table~\ref{compare-table}. The results are replicated from the corresponding papers.

DARNet has outperformed the current state-of-the-art models on the KUL, DTU, and MM-AAD datasets, achieving further enhancements in performance. On the KUL dataset, the DARNet achieves average accuracies of 91.6\% (SD: 4.83\%), 96.2\% (SD: 3.04\%), 97.2\% (SD: 2.50\%) under 0.1-second, 1-second and 2-second decision window, respectively. On the DTU dataset, the DARNet achieves average accuracies of 79.5\% (SD: 5.84\%) for 0.1-second decision window, 87.8\% (SD: 6.02\%) for 1-second decision window, 89.9\% (SD: 5.03\%) for 2-second decision window, respectively. On the MM-AAD dataset, the DARNet also demonstrates outstanding decoding accuracies of 94.9\% (SD: 4.79\%) for 0.1-second, 96.0\% (SD: 4.00\%) for 1-second, 96.5\% (SD: 3.59\%) for 2-second in the audio-only scene, and 95.8\% (SD: 4.04\%) for 0.1-second, 96.4\% (SD: 3.72\%) for 1-second, 96.8\% (SD: 3.44\%) for 2-second in the audio-visual scene.

Overall, DARNet's decoding accuracy increases with larger decision windows, consistent with prior research \citep{mbssf,stanet}. This is because longer decision windows provide more information for the model to make judgments while also mitigating the impact of individual outliers on the predictions. However, DARNet still maintains excellent performance under the 0.1-second decision window. Additionally, we observe that in the MM-AAD dataset, performance is better in the audio-visual condition compared to the audio-only condition in two different scenarios. We attribute this improvement to the visual cues aiding humans in localizing sound sources.

\begin{table}[t]
    \centering
    \caption{Auditory attention detection accuracy(\%) comparison on DTU, KUL and MM-AAD dataset. The results annotated by * are taken from \citep{DBPNet}. Our experimental setup is consistent with theirs to ensure fairness in comparison. Hence, we directly cited their results.}
    \label{compare-table}
    \begin{tabular}{cccccc}
        \toprule
        \multirow{2}{*}{Dataset} & \multirow{2}{*}{Scene} & \multirow{2}{*}{Model} & \multicolumn{3}{c}{Decision Window} \\
        \cmidrule(r){4-6}
        & & & 0.1-second & 1-second & 2-second \\
        \midrule
        \multirow{6}{*}{KUL} & \multirow{6}{*}{audio-only} 
        & SSF-CNN\textsuperscript{*} \citep{cai2021low} & 76.3 \(\pm\) 8.47 & 84.4 \(\pm\) 8.67 & 87.8 \(\pm\) 7.87 \\
        & & MBSSFCC\textsuperscript{*} \citep{mbssf} & 79.0 \(\pm\) 7.34  & 86.5 \(\pm\) 7.16 & 89.5 \(\pm\) 6.74 \\  
        & & BSAnet \citep{cai2023bio} & - & 93.7 \(\pm\) 4.02 & 95.2 \(\pm\) 3.08 \\
        & & DenseNet-3D \citep{xu2024densenet} & - & 94.3 \(\pm\) 4.3 & 95.9 \(\pm\) 4.3 \\
        & & DBPNet\textsuperscript{*} \citep{DBPNet} & 87.1 \(\pm\) 6.55 & 95.0  \(\pm\) 4.16 & 96.5  \(\pm\) 3.50 \\
        & & \textbf{DARNet(ours)} & \textbf{91.6} \(\bm{\pm}\) \textbf{4.83} & \textbf{96.2} \(\bm{\pm}\) \textbf{3.04} & \textbf{97.2} \(\bm{\pm}\) \textbf{2.50} \\
        \midrule
        \multirow{6}{*}{DTU} & \multirow{6}{*}{audio-only} 
        & SSF-CNN\textsuperscript{*} \citep{cai2021low} & 62.5 \(\pm\) 3.40 & 69.8 \(\pm\) 5.12 & 73.3 \(\pm\) 6.21 \\
        & & MBSSFCC\textsuperscript{*} \citep{mbssf} & 66.9 \(\pm\) 5.00 & 75.6 \(\pm\) 6.55 & 78.7 \(\pm\) 6.75 \\  
        & & BSAnet \citep{cai2023bio} & - & 83.1 \(\pm\) 6.75 & 85.6 \(\pm\) 6.47 \\
        & & EEG-Graph Net \citep{cai2023brain} & 72.5 \(\pm\) 7.41 & 78.7 \(\pm\) 6.47 & 79.4 \(\pm\) 7.16 \\
        & & DBPNet\textsuperscript{*} \citep{DBPNet} & 75.1  \(\pm\) 4.87 & 83.9  \(\pm\) 5.95 & 86.5  \(\pm\) 5.34 \\
        & & \textbf{DARNet(ours)} & \textbf{79.5} \(\bm{\pm}\) \textbf{5.84} & \textbf{87.8} \(\bm{\pm}\) \textbf{6.02} & \textbf{89.9} \(\bm{\pm}\) \textbf{5.03} \\
        \midrule
        \multirow{8}{*}{MM-AAD} & \multirow{4}{*}{audio-only} 
        & SSF-CNN\textsuperscript{*} \citep{cai2021low} & 56.5 \(\pm\) 5.71 & 57.0 \(\pm\) 6.55 & 57.9 \(\pm\) 7.47 \\
        & & MBSSFCC\textsuperscript{*} \citep{mbssf} & 75.3 \(\pm\) 9.27 & 76.5 \(\pm\) 9.90 & 77.0 \(\pm\) 9.92 \\
        & & DBPNet\textsuperscript{*} \citep{DBPNet} & 91.4 \(\pm\) 4.63 & 92.0  \(\pm\) 5.42 & 92.5  \(\pm\) 4.59 \\
        & & \textbf{DARNet(ours)} & \textbf{94.9} \(\bm{\pm}\) \textbf{4.79} & \textbf{96.0} \(\bm{\pm}\) \textbf{4.00} & \textbf{96.5} \(\bm{\pm}\) \textbf{3.59} \\
        \cmidrule(r){2-6}
        & \multirow{4}{*}{audio-visual} 
        & SSF-CNN\textsuperscript{*} \citep{cai2021low} & 56.6 \(\pm\) 3.82 & 57.2 \(\pm\) 5.59 & 58.2 \(\pm\) 6.39 \\
        & & MBSSFCC\textsuperscript{*} \citep{mbssf} & 77.2 \(\pm\) 9.01 & 78.1 \(\pm\) 10.1 & 78.4 \(\pm\) 9.57 \\        
        & & DBPNet\textsuperscript{*} \citep{DBPNet} & 92.1 \(\pm\) 4.47 & 92.8 \(\pm\) 5.94 & 93.4 \(\pm\) 4.86 \\
        & & \textbf{DARNet(ours)} & \textbf{95.8} \(\bm{\pm}\) \textbf{4.04} & \textbf{96.4} \(\bm{\pm}\) \textbf{3.72} & \textbf{96.8} \(\bm{\pm}\) \textbf{3.44} \\
        \bottomrule
    \end{tabular}
    
\end{table}

\subsection{Ablation Study}
We conducted comprehensive ablation experiments by removing the spatial feature extraction module, the temporal feature extraction module, and the feature fusion module. Additionally, we supplemented our study with ablation experiments using a single-layer attention refinement module on the KUL and DTU dataset, referred to as single-DARNet. All experimental conditions remained the same as in previous settings. Additionally, we ensured that all model network parameters were fully optimized to guarantee that the model's performance reached its best under each condition, whether a module was removed or added. The results of the ablation experiments are shown in Table~\ref{ablation-table}.

Experimental results show that on the DTU dataset, after removing the spatial feature extraction module from DARNet, the average accuracy decreased by 10.1\% under a 0.1s decision window, 13.1\% under a 1s decision window, and 12.0\% under a 2s decision window. After removing the temporal feature extraction module, the average accuracy for the 0.1s, 1s, and 2s decision windows decreased by 10.9\%, 14.2\%, and 11.9\%, respectively. After removing the feature fusion module, the average accuracy decreased by 2.5\% under a 0.1s decision window, 1.6\% under a 1s decision window, and 1.4\% under a 2s decision window. On the KUL dataset and the MM-AAD dataset, removing the aforementioned modules also resulted in similar trends of decreased average accuracy.

\begin{table}
    \centering
    \caption{Ablation Study on KUL, DTU, and MM-AAD dataset.}
    \label{ablation-table}
    \begin{tabular}{cccccc}
        \toprule
        \multirow{2}{*}{Dataset} & \multirow{2}{*}{Scene} & \multirow{2}{*}{Model} & \multicolumn{3}{c}{Decision Window} \\
        \cmidrule(r){4-6}
        & & & 0.1-second & 1-second & 2-second \\
        \midrule
        \multirow{5}{*}{KUL} & \multirow{5}{*}{audio-only} 
        & w/o spatial feature & 81.1 \(\pm\) 6.51 & 87.5 \(\pm\) 7.24 & 89.0  \(\pm\) 4.93 \\
        & & w/o temporal feature & 81.2 \(\pm\) 6.94 & 86.7 \(\pm\) 7.30 & 89.1 \(\pm\) 5.37 \\
        & & w/o feature fusion & 90.5 \(\pm\) 5.09 & 95.2  \(\pm\) 3.58 & 96.1  \(\pm\) 3.46 \\
        & & single-DARNet & 91.1 \(\pm\) 5.18 & 95.5 \(\pm\) 3.28 & 96.2 \(\pm\) 2.98 \\
        & & \textbf{DARNet(ours)} & \textbf{91.6} \(\bm{\pm}\) \textbf{4.83} & \textbf{96.2} \(\bm{\pm}\) \textbf{3.04} & \textbf{97.2} \(\bm{\pm}\) \textbf{2.50} \\
        \midrule
        \multirow{5}{*}{DTU} & \multirow{5}{*}{audio-only} 
        & w/o spatial feature & 71.5 \(\pm\) 5.79 & 76.3 \(\pm\) 7.21 & 79.1 \(\pm\) 6.79 \\
        & &  w/o temporal feature & 70.9 \(\pm\) 5.62 & 75.3 \(\pm\) 6.62 & 79.2 \(\pm\) 6.84 \\
        & & w/o feature fusion & 77.5 \(\pm\) 7.07 & 86.3 \(\pm\) 6.16 & 88.6 \(\pm\) 5.71 \\
        & & single-DARNet & 79.1 \(\pm\) 5.66 & 86.3 \(\pm\) 5.83 & 88.0 \(\pm\) 4.03 \\
        & & \textbf{DARNet(ours)} & \textbf{79.5} \(\bm{\pm}\) \textbf{5.84} & \textbf{87.8} \(\bm{\pm}\) \textbf{6.02} & \textbf{89.9} \(\bm{\pm}\) \textbf{5.03} \\
        \midrule
        \multirow{8}{*}{MM-AAD} & \multirow{4}{*}{audio-only} 
        & w/o spatial feature & 90.0 \(\pm\) 5.76 & 91.0 \(\pm\) 5.38 & 92.5 \(\pm\) 4.76 \\
        & &  w/o temporal feature & 87.8 \(\pm\) 5.66 & 89.0 \(\pm\) 5.20 & 91.1 \(\pm\) 5.31 \\
        & & w/o feature fusion & 94.3 \(\pm\) 3.94 & 94.8  \(\pm\) 4.05 & 95.7  \(\pm\) 4.21 \\
        & & \textbf{DARNet(ours)} & \textbf{94.9} \(\bm{\pm}\) \textbf{4.79} & \textbf{96.0} \(\bm{\pm}\) \textbf{4.00} & \textbf{96.5} \(\bm{\pm}\) \textbf{3.59} \\
        \cmidrule(r){2-6}
        & \multirow{4}{*}{audio-visual} 
        & w/o spatial feature & 90.4 \(\pm\) 5.99 & 91.2 \(\pm\) 5.56 & 93.1  \(\pm\) 5.36 \\
        & &  w/o temporal feature & 89.4 \(\pm\) 6.96 & 90.5 \(\pm\) 6.04 & 92.1 \(\pm\) 5.72 \\
        & & w/o feature fusion & 95.3 \(\pm\) 4.57 & 95.7  \(\pm\) 3.88 & 96.1  \(\pm\) 3.96 \\
        & & \textbf{DARNet(ours)} & \textbf{95.8} \(\bm{\pm}\) \textbf{4.04} & \textbf{96.4} \(\bm{\pm}\) \textbf{3.72} & \textbf{96.8} \(\bm{\pm}\) \textbf{3.44} \\
        \bottomrule
    \end{tabular}
\end{table}

\subsection{Experimental Correction}
To ensure fairness in comparisons, we aligned our previous experimental setup and data processing methods with those of DBPNet. However, during the code review for the final version of this paper, we found that DBPNet applies CSP to the data prior to dataset splitting, which can lead to data leakage. Consequently, we corrected the processing steps and conducted additional experiments, as shown in Table~\ref{Experimental Correction}. The data preprocessing steps for the supplementary experiments are as follows:  
\begin{enumerate}
    \item Split each trial of the dataset into the first 90\% for training and the last 10\% for testing.
    \item Fit the CSP transformation matrix using the training data, project the data to extract features, and apply the matrix to the testing data.
    \item Apply a sliding window with 50\% overlap to the processed training data, then randomly split into a training (90\%) and a validation (10\%) set. Similarly, apply a sliding window to the  processed testing data as the final test set.
\end{enumerate}

Following these steps, we rigorously avoided risk of data leakage. Experimental results show that both corrected DBPNet\(^{2}\) and DARNet\(^{2}\) exhibit a performance decline compared to the original DBPNet\(^{1}\) and DARNet\(^{1}\). However, DARNet\(^{2}\) still achieves state-of-the-art performance under most test conditions, as shown in Table~\ref{Experimental Correction}. For instance, on the KUL dataset with a 0.1-second decision window, DARNet\(^{2}\) maintains a decoding accuracy of 89.2\%.

\begin{table}[h]
    \centering
    \caption{Experimental Correction on KUL, DTU, and MM-AAD dataset. The experimental setup for the results marked with \(^{1}\) is consistent with DBPNet\citep{DBPNet}, where CSP is applied prior to dataset splitting. In contrast, for the results marked with \(^{2}\), CSP is applied after dataset splitting.}
    \label{Experimental Correction}
    \begin{tabular}{cccccc}
        \toprule
        \multirow{2}{*}{Dataset} & \multirow{2}{*}{Scene} & \multirow{2}{*}{Model} & \multicolumn{3}{c}{Decision Window} \\
        \cmidrule(r){4-6}
        & & & 0.1-second & 1-second & 2-second \\
        \midrule
        \multirow{4}{*}{KUL} & \multirow{4}{*}{audio-only} 
        & DBPNet\(^{1}\) & 87.1 \(\pm\) 6.55 & 95.0 \(\pm\) 4.16 & 96.5  \(\pm\) 3.50 \\
        & & DBPNet\(^{2}\) & 85.3 \(\pm\) 6.22 & 94.4 \(\pm\) 4.62 & 95.3 \(\pm\) 4.63 \\
        & & DARNet\(^{1}\) & 91.6 \(\pm\) 4.83 & 96.2 \(\pm\) 3.04 & 97.2 \(\pm\) 2.50 \\
        & & DARNet\(^{2}\) & 89.2 \(\pm\) 5.50 & 94.8  \(\pm\) 4.53 & 95.5  \(\pm\) 4.89 \\
        \midrule
        \multirow{4}{*}{DTU} & \multirow{4}{*}{audio-only} 
        & DBPNet\(^{1}\) & 75.1  \(\pm\) 4.87 & 83.9  \(\pm\) 5.95 & 86.5  \(\pm\) 5.34 \\
        & & DBPNet\(^{2}\) & 74.0 \(\pm\) 5.20 & 79.8 \(\pm\) 6.91 & 80.2 \(\pm\) 6.79 \\
        & & DARNet\(^{1}\) & 79.5 \(\pm\) 5.84 & 87.8 \(\pm\) 6.02 & 89.9 \(\pm\) 5.03 \\
        & & DARNet\(^{2}\) & 74.6 \(\pm\) 6.09 & 80.1 \(\pm\) 6.85 & 81.2  \(\pm\) 6.34 \\
        \midrule
        \multirow{8}{*}{MM-AAD} & \multirow{4}{*}{audio-only} 
        & DBPNet\(^{1}\) & 91.4 \(\pm\) 4.63 & 92.0  \(\pm\) 5.42 & 92.5  \(\pm\) 4.59 \\
        & & DBPNet\(^{2}\) & 90.0 \(\pm\) 5.51 & 90.7 \(\pm\) 5.68 & 91.6 \(\pm\) 4.82 \\
        & & DARNet\(^{1}\) & 94.9  \(\pm\) 4.79 & 96.0  \(\pm\) 4.00 & 96.5  \(\pm\) 3.59 \\
        & & DARNet\(^{2}\) & 91.5 \(\pm\) 5.27 & 92.2  \(\pm\) 4.54 & 92.8  \(\pm\) 5.22 \\
        \cmidrule(r){2-6}
        & \multirow{4}{*}{audio-visual} 
        & DBPNet\(^{1}\) & 92.1 \(\pm\) 4.47 & 92.8 \(\pm\) 5.94 & 93.4 \(\pm\) 4.86 \\
        & & DBPNet\(^{2}\) & 92.0 \(\pm\) 5.51 & 93.0 \(\pm\) 5.19 & 92.7 \(\pm\) 6.04 \\
        & & DARNet\(^{1}\) & 95.8 \(\pm\) 4.04 & 96.4\(\pm\) 3.72 & 96.8 \(\pm\) 3.44 \\
        & & DARNet\(^{2}\) & 92.7 \(\pm\) 5.34 & 93.4\(\pm\) 5.23 & 94.2 \(\pm\) 4.84 \\
        \bottomrule
    \end{tabular}
\end{table}

\section{Discussion}
\subsection{Comparative Analysis}
To further evaluate the performance of our proposed DARNet, we compared it with other advanced AAD models, as shown in Table~\ref{compare-table}. The results indicate that our DARNet has achieved a significant improvement over the current state-of-the-art results.

For example, on the DTU dataset, our DARNet has shown relative improvements of 27.2\%, 18.8\%, 9.7\%, and 5.9\% for 0.1-second decision window, compared to the SSF-CNN, MBSSFCC, EEG-Graph Net and DBPNet models, respectively. Compared to the SSF-CNN, MBSSFCC, BSAnet, EEG-Graph Net, and DBPNet models, the relative improvements achieve 25.8\%, 16.1\%, 5.7\%, 11.6\%, 4.6\% for 1-second decision window, and 22.6\%, 14.2\%, 5.0\%, 13.2\%, 3.9\% for 2-second. On both the KUL and MM-AAD datasets, DARNet has achieved similar improvements compared to the state-of-the-art models. The particularly outstanding results achieved across all three datasets under the 0.1-second decision window indicate the potential of DARNet for real-time decoding of auditory attention.

Overall, the excellent performance of DARNet across different datasets and decision windows demonstrates its robustness and versatility in various contexts. This further validates the potential of DARNet as an effective EEG analysis model and provides strong support for its widespread application in real-world scenarios.

\subsection{Ablation Analysis}
As shown in Table~\ref{ablation-table} and Figure~\ref{ablation_figure}, compared with removing the spatial feature extraction step, removing the temporal feature extraction step, removing the feature fusion module, and using a single-layer attention refinement module, we believe DARNet performs excellently for the following reasons:

\textbf{1. Integrating multiple sources of information:} DARNet integrates temporal and spatial distribution features from EEG signals, constructing richer and more robust spatiotemporal features. This enables the model to comprehensively understand the spatiotemporal information within EEG signals, thereby enhancing the understanding of brain activity. In contrast, removing any single feature may lead to information loss or the inability to capture the transient changes in EEG signals, thereby impacting the model's performance.
    
\textbf{2. Comprehensive capture of temporal dependencies:} The dual attention refinement module and feature fusion module of DARNet comprehensively capture temporal patterns and dependencies at different levels, enabling the model to better understand the temporal dynamics within EEG signals. This holistic consideration of features at different time scales is crucial for the analysis of EEG data.

\textbf{3. Robust feature representation:} Despite observing that removing the feature fusion module did not lead to a significant decrease in accuracy across the three datasets, the performance variability of DARNet increases substantially. We believe that the feature fusion module integrates temporal patterns and dependencies at different levels, enabling the model to better understand and utilize the complex relationships within the data, thus enhancing the robustness and generalization of the model.

\begin{table}[h]
    \centering
    \caption{The training parameter counts comparison. "M" denotes a million.}
    \begin{tabular}{cc}
        \toprule
        Model & Trainable Parameters Counts\\
        \midrule
         SSF-CNN \citep{cai2021low} & 4.21M \\
         MBSSFCC \citep{mbssf} & 83.91M \\
         DBPNet \citep{DBPNet} & 0.91M \\
         \textbf{DARNet (ours)} & \textbf{0.08M} \\
        \bottomrule
    \end{tabular}

    \label{parameters-table}
\end{table}

\subsection{Computational Cost}
We compare the training parameter counts of our DARNet, SSF-CNN \citep{cai2021low}, MBSSFCC \citep{mbssf}, and DBPNet \citep{DBPNet}, with the results shown in Table~\ref{parameters-table}. The parameter count of DARNet is 51.6 times lower than that of SSF-CNN, 1331.5 times lower than that of MBSSFCC, and 10.4 times lower than that of DBPNet. Compared to other models, DARNet demonstrates superior parameter efficiency. Despite having fewer parameters, DARNet maintains good performance, indicating its ability to be applied in resource-constrained environments for AAD analysis, thus demonstrating practical utility.

\section{Conclusion}
In this paper, we propose the DARNet, a novel dual attention refinement network with spatiotemporal construction for auditory attention detection. By employing spatial convolution operations across all channels, DARNet effectively leverages the spatial information embedded in EEG signals, thereby constructing a more robust spatiotemporal feature. Additionally, DARNet integrates dual attention refinement and feature fusion techniques to comprehensively capture temporal patterns and dependencies at various levels, enhancing the model's ability to capture the temporal dynamics within EEG signals. We evaluate the performance of DARNet on three datasets: KUL, DTU, and MM-AAD. DARNet achieves a decoding accuracy of 96.2\% on the 1-second decision window of the KUL dataset and 87.8\% on the 1-second decision window of the DTU dataset, demonstrating significant improvements compared to current state-of-the-art models. The experimental results validate the effectiveness and efficiency of the DARNet architecture, indicating its potential for practical applications. In future research, we plan to further explore DARNet's performance on cross-subject tasks to verify its generalization and robustness.

\section*{Acknowledgments and Disclosure of Funding}
This work is supported by the {STI 2030—Major Projects (No. 2021ZD0201500)}, the National Natural Science Foundation of China (NSFC) (No.62201002, 6247077204), Excellent Youth Foundation of Anhui Scientific Committee (No. 2408085Y034), Distinguished Youth Foundation of Anhui Scientific Committee (No. 2208085J05), Special Fund for Key Program of Science and Technology of Anhui Province (No. 202203a07020008), Open Fund of Key Laboratory of Flight Techniques and Flight Safety, CACC (No, FZ2022KF15), Cloud Ginger XR-1.

\bibliographystyle{unsrtnat}
\bibliography{DARNet}

\newpage
\section*{NeurIPS Paper Checklist}

\begin{enumerate}

\item {\bf Claims}
    \item[] Question: Do the main claims made in the abstract and introduction accurately reflect the paper's contributions and scope?
    \item[] Answer: \answerYes{} 
    \item[] Justification: The abstract and introduction of the paper clearly state the main contributions and scope of the research. The claims made are aligned with the theoretical and experimental results presented in the paper. 
    \item[] Guidelines:
    \begin{itemize}
        \item The answer NA means that the abstract and introduction do not include the claims made in the paper.
        \item The abstract and/or introduction should clearly state the claims made, including the contributions made in the paper and important assumptions and limitations. A No or NA answer to this question will not be perceived well by the reviewers. 
        \item The claims made should match theoretical and experimental results, and reflect how much the results can be expected to generalize to other settings. 
        \item It is fine to include aspirational goals as motivation as long as it is clear that these goals are not attained by the paper. 
    \end{itemize}

\item {\bf Limitations}
    \item[] Question: Does the paper discuss the limitations of the work performed by the authors?
    \item[] Answer: \answerYes{} 
    \item[] Justification: We outlined the limitations in the conclusion. Current AAD research primarily employs two experimental strategies: subject-dependent and subject-independent.Subject-dependent refers to the training and evaluating procedures containing only samples from a single subject, while subject-independent contains samples from all subjects in the dataset. Our proposed model has been validated under the subject-dependent condition and has demonstrated exceptional results. However, further exploration and resolution of the issue of inter-subject variability are necessary to enable our model to be more widely applicable to real-world brain-computer interface applications.
    \item[] Guidelines:
    \begin{itemize}
        \item The answer NA means that the paper has no limitation while the answer No means that the paper has limitations, but those are not discussed in the paper. 
        \item The authors are encouraged to create a separate "Limitations" section in their paper.
        \item The paper should point out any strong assumptions and how robust the results are to violations of these assumptions (e.g., independence assumptions, noiseless settings, model well-specification, asymptotic approximations only holding locally). The authors should reflect on how these assumptions might be violated in practice and what the implications would be.
        \item The authors should reflect on the scope of the claims made, e.g., if the approach was only tested on a few datasets or with a few runs. In general, empirical results often depend on implicit assumptions, which should be articulated.
        \item The authors should reflect on the factors that influence the performance of the approach. For example, a facial recognition algorithm may perform poorly when image resolution is low or images are taken in low lighting. Or a speech-to-text system might not be used reliably to provide closed captions for online lectures because it fails to handle technical jargon.
        \item The authors should discuss the computational efficiency of the proposed algorithms and how they scale with dataset size.
        \item If applicable, the authors should discuss possible limitations of their approach to address problems of privacy and fairness.
        \item While the authors might fear that complete honesty about limitations might be used by reviewers as grounds for rejection, a worse outcome might be that reviewers discover limitations that aren't acknowledged in the paper. The authors should use their best judgment and recognize that individual actions in favor of transparency play an important role in developing norms that preserve the integrity of the community. Reviewers will be specifically instructed to not penalize honesty concerning limitations.
    \end{itemize}

\item {\bf Theory Assumptions and Proofs}
    \item[] Question: For each theoretical result, does the paper provide the full set of assumptions and a complete (and correct) proof?
    \item[] Answer: \answerYes{} 
    \item[] Justification: All the theorems, formulas, and proofs in the paper have been properly numbered and cross-referenced, fulfilling the guidelines provided.
    \item[] Guidelines:
    \begin{itemize}
        \item The answer NA means that the paper does not include theoretical results. 
        \item All the theorems, formulas, and proofs in the paper should be numbered and cross-referenced.
        \item All assumptions should be clearly stated or referenced in the statement of any theorems.
        \item The proofs can either appear in the main paper or the supplemental material, but if they appear in the supplemental material, the authors are encouraged to provide a short proof sketch to provide intuition. 
        \item Inversely, any informal proof provided in the core of the paper should be complemented by formal proofs provided in appendix or supplemental material.
        \item Theorems and Lemmas that the proof relies upon should be properly referenced. 
    \end{itemize}

    \item {\bf Experimental Result Reproducibility}
    \item[] Question: Does the paper fully disclose all the information needed to reproduce the main experimental results of the paper to the extent that it affects the main claims and/or conclusions of the paper (regardless of whether the code and data are provided or not)?
    \item[] Answer: \answerYes{} 
    \item[] Justification: We have detailed our model and experimental setup thoroughly in the Methodology and Experiments sections, providing all necessary information to reproduce the main experimental results.
    \item[] Guidelines:
    \begin{itemize}
        \item The answer NA means that the paper does not include experiments.
        \item If the paper includes experiments, a No answer to this question will not be perceived well by the reviewers: Making the paper reproducible is important, regardless of whether the code and data are provided or not.
        \item If the contribution is a dataset and/or model, the authors should describe the steps taken to make their results reproducible or verifiable. 
        \item Depending on the contribution, reproducibility can be accomplished in various ways. For example, if the contribution is a novel architecture, describing the architecture fully might suffice, or if the contribution is a specific model and empirical evaluation, it may be necessary to either make it possible for others to replicate the model with the same dataset, or provide access to the model. In general. releasing code and data is often one good way to accomplish this, but reproducibility can also be provided via detailed instructions for how to replicate the results, access to a hosted model (e.g., in the case of a large language model), releasing of a model checkpoint, or other means that are appropriate to the research performed.
        \item While NeurIPS does not require releasing code, the conference does require all submissions to provide some reasonable avenue for reproducibility, which may depend on the nature of the contribution. For example
        \begin{enumerate}
            \item If the contribution is primarily a new algorithm, the paper should make it clear how to reproduce that algorithm.
            \item If the contribution is primarily a new model architecture, the paper should describe the architecture clearly and fully.
            \item If the contribution is a new model (e.g., a large language model), then there should either be a way to access this model for reproducing the results or a way to reproduce the model (e.g., with an open-source dataset or instructions for how to construct the dataset).
            \item We recognize that reproducibility may be tricky in some cases, in which case authors are welcome to describe the particular way they provide for reproducibility. In the case of closed-source models, it may be that access to the model is limited in some way (e.g., to registered users), but it should be possible for other researchers to have some path to reproducing or verifying the results.
        \end{enumerate}
    \end{itemize}

\item {\bf Open access to data and code}
    \item[] Question: Does the paper provide open access to the data and code, with sufficient instructions to faithfully reproduce the main experimental results, as described in supplemental material?
    \item[] Answer: \answerYes{} 
    \item[] Justification: The paper includes code as an attachment, facilitating the reproduction of the main experimental results.
    \item[] Guidelines:
    \begin{itemize}
        \item The answer NA means that paper does not include experiments requiring code.
        \item Please see the NeurIPS code and data submission guidelines (\url{https://nips.cc/public/guides/CodeSubmissionPolicy}) for more details.
        \item While we encourage the release of code and data, we understand that this might not be possible, so “No” is an acceptable answer. Papers cannot be rejected simply for not including code, unless this is central to the contribution (e.g., for a new open-source benchmark).
        \item The instructions should contain the exact command and environment needed to run to reproduce the results. See the NeurIPS code and data submission guidelines (\url{https://nips.cc/public/guides/CodeSubmissionPolicy}) for more details.
        \item The authors should provide instructions on data access and preparation, including how to access the raw data, preprocessed data, intermediate data, and generated data, etc.
        \item The authors should provide scripts to reproduce all experimental results for the new proposed method and baselines. If only a subset of experiments are reproducible, they should state which ones are omitted from the script and why.
        \item At submission time, to preserve anonymity, the authors should release anonymized versions (if applicable).
        \item Providing as much information as possible in supplemental material (appended to the paper) is recommended, but including URLs to data and code is permitted.
    \end{itemize}

\item {\bf Experimental Setting/Details}
    \item[] Question: Does the paper specify all the training and test details (e.g., data splits, hyperparameters, how they were chosen, type of optimizer, etc.) necessary to understand the results?
    \item[] Answer: \answerYes{} 
    \item[] Justification: We have provided detailed specifications of our experimental settings in the "Methodology" and "Experiments" sections of the paper. This includes descriptions of the data splits, hyperparameters, selection criteria, and the type of optimizer used. The comprehensive documentation of these parameters ensures that our results can be understood and replicated by other researchers.
    \item[] Guidelines:
    \begin{itemize}
        \item The answer NA means that the paper does not include experiments.
        \item The experimental setting should be presented in the core of the paper to a level of detail that is necessary to appreciate the results and make sense of them.
        \item The full details can be provided either with the code, in appendix, or as supplemental material.
    \end{itemize}

\item {\bf Experiment Statistical Significance}
    \item[] Question: Does the paper report error bars suitably and correctly defined or other appropriate information about the statistical significance of the experiments?
    \item[] Answer: \answerNo{} 
    \item[] Justification: The experimental results do not include confidence intervals or statistical significance tests.
    \item[] Guidelines:
    \begin{itemize}
        \item The answer NA means that the paper does not include experiments.
        \item The authors should answer "Yes" if the results are accompanied by error bars, confidence intervals, or statistical significance tests, at least for the experiments that support the main claims of the paper.
        \item The factors of variability that the error bars are capturing should be clearly stated (for example, train/test split, initialization, random drawing of some parameter, or overall run with given experimental conditions).
        \item The method for calculating the error bars should be explained (closed form formula, call to a library function, bootstrap, etc.)
        \item The assumptions made should be given (e.g., Normally distributed errors).
        \item It should be clear whether the error bar is the standard deviation or the standard error of the mean.
        \item It is OK to report 1-sigma error bars, but one should state it. The authors should preferably report a 2-sigma error bar than state that they have a 96\% CI, if the hypothesis of Normality of errors is not verified.
        \item For asymmetric distributions, the authors should be careful not to show in tables or figures symmetric error bars that would yield results that are out of range (e.g. negative error rates).
        \item If error bars are reported in tables or plots, The authors should explain in the text how they were calculated and reference the corresponding figures or tables in the text.
    \end{itemize}

\item {\bf Experiments Compute Resources}
    \item[] Question: For each experiment, does the paper provide sufficient information on the computer resources (type of compute workers, memory, time of execution) needed to reproduce the experiments?
    \item[] Answer: \answerNo{} 
    \item[] Justification: Our model contains only 0.78 million training parameters, making it lightweight and capable of running on most machines. Therefore, we did not provide detailed specifications of the compute resources required.
    \item[] Guidelines:
    \begin{itemize}
        \item The answer NA means that the paper does not include experiments.
        \item The paper should indicate the type of compute workers CPU or GPU, internal cluster, or cloud provider, including relevant memory and storage.
        \item The paper should provide the amount of compute required for each of the individual experimental runs as well as estimate the total compute. 
        \item The paper should disclose whether the full research project required more compute than the experiments reported in the paper (e.g., preliminary or failed experiments that didn't make it into the paper). 
    \end{itemize}
    
\item {\bf Code Of Ethics}
    \item[] Question: Does the research conducted in the paper conform, in every respect, with the NeurIPS Code of Ethics \url{https://neurips.cc/public/EthicsGuidelines}?
    \item[] Answer: \answerYes{} 
    \item[] Justification: Yes, our research adheres to all ethical guidelines required by NeurIPS.
    \item[] Guidelines:
    \begin{itemize}
        \item The answer NA means that the authors have not reviewed the NeurIPS Code of Ethics.
        \item If the authors answer No, they should explain the special circumstances that require a deviation from the Code of Ethics.
        \item The authors should make sure to preserve anonymity (e.g., if there is a special consideration due to laws or regulations in their jurisdiction).
    \end{itemize}

\item {\bf Broader Impacts}
    \item[] Question: Does the paper discuss both potential positive societal impacts and negative societal impacts of the work performed?
    \item[] Answer: \answerNA{} 
    \item[] Justification: Our research does not directly produce societal impacts as it focuses on technical advancements in a specific field without direct societal applications.
    \item[] Guidelines:
    \begin{itemize}
        \item The answer NA means that there is no societal impact of the work performed.
        \item If the authors answer NA or No, they should explain why their work has no societal impact or why the paper does not address societal impact.
        \item Examples of negative societal impacts include potential malicious or unintended uses (e.g., disinformation, generating fake profiles, surveillance), fairness considerations (e.g., deployment of technologies that could make decisions that unfairly impact specific groups), privacy considerations, and security considerations.
        \item The conference expects that many papers will be foundational research and not tied to particular applications, let alone deployments. However, if there is a direct path to any negative applications, the authors should point it out. For example, it is legitimate to point out that an improvement in the quality of generative models could be used to generate deepfakes for disinformation. On the other hand, it is not needed to point out that a generic algorithm for optimizing neural networks could enable people to train models that generate Deepfakes faster.
        \item The authors should consider possible harms that could arise when the technology is being used as intended and functioning correctly, harms that could arise when the technology is being used as intended but gives incorrect results, and harms following from (intentional or unintentional) misuse of the technology.
        \item If there are negative societal impacts, the authors could also discuss possible mitigation strategies (e.g., gated release of models, providing defenses in addition to attacks, mechanisms for monitoring misuse, mechanisms to monitor how a system learns from feedback over time, improving the efficiency and accessibility of ML).
    \end{itemize}
    
\item {\bf Safeguards}
    \item[] Question: Does the paper describe safeguards that have been put in place for responsible release of data or models that have a high risk for misuse (e.g., pretrained language models, image generators, or scraped datasets)?
    \item[] Answer: \answerNA{} 
    \item[] Justification: This paper poses no such risks.
    \item[] Guidelines:
    \begin{itemize}
        \item The answer NA means that the paper poses no such risks.
        \item Released models that have a high risk for misuse or dual-use should be released with necessary safeguards to allow for controlled use of the model, for example by requiring that users adhere to usage guidelines or restrictions to access the model or implementing safety filters. 
        \item Datasets that have been scraped from the Internet could pose safety risks. The authors should describe how they avoided releasing unsafe images.
        \item We recognize that providing effective safeguards is challenging, and many papers do not require this, but we encourage authors to take this into account and make a best faith effort.
    \end{itemize}

\item {\bf Licenses for existing assets}
    \item[] Question: Are the creators or original owners of assets (e.g., code, data, models), used in the paper, properly credited and are the license and terms of use explicitly mentioned and properly respected?
    \item[] Answer: \answerYes{} 
    \item[] Justification: All datasets used in our paper are publicly available datasets, and we have cited the respective literature for each dataset. Any researcher can download these datasets from the provided sources.
    \item[] Guidelines:
    \begin{itemize}
        \item The answer NA means that the paper does not use existing assets.
        \item The authors should cite the original paper that produced the code package or dataset.
        \item The authors should state which version of the asset is used and, if possible, include a URL.
        \item The name of the license (e.g., CC-BY 4.0) should be included for each asset.
        \item For scraped data from a particular source (e.g., website), the copyright and terms of service of that source should be provided.
        \item If assets are released, the license, copyright information, and terms of use in the package should be provided. For popular datasets, \url{paperswithcode.com/datasets} has curated licenses for some datasets. Their licensing guide can help determine the license of a dataset.
        \item For existing datasets that are re-packaged, both the original license and the license of the derived asset (if it has changed) should be provided.
        \item If this information is not available online, the authors are encouraged to reach out to the asset's creators.
    \end{itemize}

\item {\bf New Assets}
    \item[] Question: Are new assets introduced in the paper well documented and is the documentation provided alongside the assets?
    \item[] Answer: \answerYes{} 
    \item[] Justification: The paper includes the submission of the model's source code.
    \item[] Guidelines:
    \begin{itemize}
        \item The answer NA means that the paper does not release new assets.
        \item Researchers should communicate the details of the dataset/code/model as part of their submissions via structured templates. This includes details about training, license, limitations, etc. 
        \item The paper should discuss whether and how consent was obtained from people whose asset is used.
        \item At submission time, remember to anonymize your assets (if applicable). You can either create an anonymized URL or include an anonymized zip file.
    \end{itemize}

\item {\bf Crowdsourcing and Research with Human Subjects}
    \item[] Question: For crowdsourcing experiments and research with human subjects, does the paper include the full text of instructions given to participants and screenshots, if applicable, as well as details about compensation (if any)? 
    \item[] Answer: \answerNA{} 
    \item[] Justification: The paper does not involve crowdsourcing nor research with human subjects.
    \item[] Guidelines:
    \begin{itemize}
        \item The answer NA means that the paper does not involve crowdsourcing nor research with human subjects.
        \item Including this information in the supplemental material is fine, but if the main contribution of the paper involves human subjects, then as much detail as possible should be included in the main paper. 
        \item According to the NeurIPS Code of Ethics, workers involved in data collection, curation, or other labor should be paid at least the minimum wage in the country of the data collector. 
    \end{itemize}

\item {\bf Institutional Review Board (IRB) Approvals or Equivalent for Research with Human Subjects}
    \item[] Question: Does the paper describe potential risks incurred by study participants, whether such risks were disclosed to the subjects, and whether Institutional Review Board (IRB) approvals (or an equivalent approval/review based on the requirements of your country or institution) were obtained?
    \item[] Answer: \answerNA{} 
    \item[] Justification: The paper does not involve crowdsourcing nor research with human subjects.
    \item[] Guidelines:
    \begin{itemize}
        \item The answer NA means that the paper does not involve crowdsourcing nor research with human subjects.
        \item Depending on the country in which research is conducted, IRB approval (or equivalent) may be required for any human subjects research. If you obtained IRB approval, you should clearly state this in the paper. 
        \item We recognize that the procedures for this may vary significantly between institutions and locations, and we expect authors to adhere to the NeurIPS Code of Ethics and the guidelines for their institution. 
        \item For initial submissions, do not include any information that would break anonymity (if applicable), such as the institution conducting the review.
    \end{itemize}

\end{enumerate}

\end{document}